\documentclass[a4paper, 11pt]{article}

\usepackage[english]{babel}
\usepackage[utf8x]{inputenc}
\usepackage[normalem]{ulem}
\usepackage[T1]{fontenc}

\usepackage{subfiles}
\usepackage{amsmath,amsthm,amsfonts,amssymb,bbm,algorithm,algorithmic}
\usepackage{rotating}
\usepackage[dvipsnames]{xcolor}
\usepackage[semicolon]{natbib} 
\usepackage{xr-hyper}
\usepackage[pdftex,                %
bookmarks         = true,
bookmarksnumbered = true,
pdfpagemode       = None,
pdfstartview      = FitH,
pdfpagelayout     = SinglePage,
colorlinks        = true,
linkcolor		  = magenta,
citecolor		  = blue,
urlcolor          = orange,
pdfborder         = {0 0 0}
]{hyperref}
\usepackage[font=small,labelfont=bf,tableposition=top]{caption}
\usepackage{graphicx,subcaption,soul,makecell,multirow,booktabs,url}

\usepackage[top=3cm, bottom=3cm, left=2.5cm, right=2cm]{geometry}

\usepackage{newtxtext}
\usepackage[subscriptcorrection]{newtxmath}

\DeclareMathOperator{\R}{\mathbb{R}}

\DeclareMathOperator{\ind}{\mathbbm{1}}
\renewcommand{\P}{\mathbb{P}}
\newcommand{\indep}{\perp \!\!\! \perp}

\newcommand{\EM}{\ensuremath}

\newcommand{\cD}{\EM{\mathcal{D}}}

\newcommand{\cH}{\EM{\mathcal{H}}}

\newcommand{\cN}{\EM{\mathcal{N}}}

\newcommand{\cU}{\EM{\mathcal{U}}}

\newcommand{\cX}{\EM{\mathcal{X}}}
\newcommand{\cY}{\EM{\mathcal{Y}}}

\newtheorem{theorem}{Theorem}
\newtheorem{lemma}[theorem]{Lemma}

\theoremstyle{definition}

\theoremstyle{remark}
\newtheorem{remark}{Remark}

\DeclareMathOperator{\Dcal}{\cD_\text{cal}}
\DeclareMathOperator{\Dtrain}{\cD_\text{train}}

\begin{document}

\begin{center}
{\LARGE
	{Two-sided conformalized survival analysis}
}

\bigskip
C. Holmes$^{1}$ and A. Marandon$^{2}\footnotemark{}$
\bigskip

{\small
{
$^{1}$ Department of Statistics, University of Oxford\\
$^{2}$ The Alan Turing Institute 
}
}
\bigskip

\end{center}

\footnotetext{Corresponding author: amarandon-carlhian@turing.ac.uk}

\begin{abstract}
This paper presents a conformal prediction procedure to generate two-sided or one-sided prediction intervals for survival times in the presence of right censoring. Specifically, the method provides two-sided predictive bounds for individuals deemed sufficiently similar to the uncensored population, while returning a lower predictive bound for others. The prediction intervals offer finite-sample coverage guarantees, requiring no distributional assumptions other than the sampled data points are independent and identically distributed. The 
performance and validity of the procedure is evaluated on both synthetic and real-world datasets.
\end{abstract}


\section{Introduction}

Prediction of survival times is a fundamental task in healthcare applications, for example, following a diagnosis or treatment. A main challenge is that survival data is generally censored, which means that the survival time is missing for some patients: either a lower bound (right-censoring) and/or an upper bound (left-censoring) on the survival time is observed instead of the true value. This work focuses on the case of right-censoring, which is the most common scenario encountered in practice. For example, if the event is death, a patient may be lost to follow-up, die from another cause other than the disease of interest, dropout from the study (for reasons related or not to disease progression), or still be alive at the time of data collection \citep{leung1997censoring}.  
There exists a breadth of methods for modelling survival \citep{tutz2016modeling, introbook2012}. Classical techniques include the Kaplan-Meier estimator \citep{kaplan1958nonparametric}, Cox's proportional hazards model \citep{cox1972regression}, the AFT model \citep{wei1992accelerated}, or Random Survival Forest  \citep{ishwaran08}. Recent years have also seen the development of deep-learning-based methods \citep{katzman2018deepsurv, zhong2021deep,wiegrebe2024deep}. Survival methods generally estimate the survival function, from which a prediction for the survival time can be deduced. 
However, as any predictive model cannot be perfect, none of these methods can provide the guarantee that single-point predictions are correct. 
Hence, accurate uncertainty estimates in the predictions \citep{banerji23} are important in such a high-stakes context, where mistakes are costly.
To this end, we consider the task of generating a \textit{prediction interval}, i.e. a lower bound and an upper bound on the survival time, which holds with a pre-fixed level of confidence.

Let $X, T, C$ be respectively the covariate vector, the survival time, and the censoring time. Define $\tilde T=\min(T, C)$ the minimum between the survival time and the censoring time as the \textit{censored survival time} and $\Delta = \ind \{ T \leq C\}$ as the \textit{event}. Formally, letting $(X_i, T_i, C_i)$ be $n+1$ independent and identically distributed (i.i.d.) copies of $X, T, C$, we observe $(X_i, \tilde T_i, \Delta_i)_{i=1}^{n}$ and $X_{n+1}$, where $X_{n+1}$ represents a new patient for which the unknown survival time $T_{n+1}$ is of interest.
The aim is to build a prediction set $\hat C_\alpha(X_{n+1}) \subset \R^{+}$ containing the survival time $T_{n+1}$ of the new patient $n+1$ with probability above a level $1-\alpha$, that is:
\begin{align*} 
 \P( T_{n+1} \in \hat C_\alpha(X_{n+1}) ) \geq 1-\alpha ,
 \end{align*}
where in the above equation, the probability is over $(X_i, \tilde T_i, \Delta_i)_{i=1}^n$ and $(X_{n+1}, T_{n+1})$.

In a general regression or classification task, prediction sets which cover the true outcome/label with a fixed probability can be constructed with conformal prediction (CP) \citep{vovk2005algorithmic, papadopoulos2002inductive, angelopoulos2021gentle, lei2018distribution,romano2019conformalized}. 
In a nutshell, CP uses ground truth examples to rigorously calibrate the predictions of arbitrary predictive models, including black-box ML algorithms.
Specifically, \textit{Split}-CP relies on a non-conformity score function $\nu(x,y)$, measuring the inadequacy of an outcome value $y$ as a prediction for an observation $x$,  based on a pre-trained predictive model. 
A prediction set is constructed by excluding outcome values with a \textit{significantly} large non-conformity score, where the threshold of exclusion is based on the distribution of non-conformity scores computed on a hold-out (\textit{calibration}) labeled dataset. 
Crucially, with CP, the coverage guarantee holds in finite sample sizes and without relying on any distributional assumptions besides exchangeability of the test and calibration data examples.

Conventional CP cannot be directly applied for predicting survival times, since, due to censoring, a sample from the true distribution of the survival times is not available. 
A naive solution is to apply CP with $\tilde T$ as the prediction target: since $\tilde T \leq T$, the lower predictive bound (LPB) is also a valid LPB for $T_{n+1}$. 
However, it is conservative, and a valid upper predictive bound (UPB) is still missing.
\cite{candes2023conformalized}
 focused on constructing a less conservative LPB on the survival time than the naive approach. Considering a setup where the censoring time is observed for every individual (censored or not), they proposed to alleviate the conservativeness of the naive method by restricting the calibration set to the censored survival times $C_i \geq c_0$, for some constant $c_0$. Under the assumption that $T \indep C \vert X$, 
 the selection $C \geq c_0$ only introduces a covariate shift with respect to the distribution of $(X, \tilde T \wedge c_0)$. Thus, the extension of conformal prediction to covariate shift \citep{tibshirani2019conformal} 
 can be used to derive a lower bound for $\tilde T_{n+1} \wedge c_0 \leq T_{n+1} \wedge c_0 \leq T_{n+1}$. 
If the covariate shift is consistently estimated, the LPB is valid asymptotically. 

\begin{figure}
    \centering
    \includegraphics[width=0.3\linewidth]{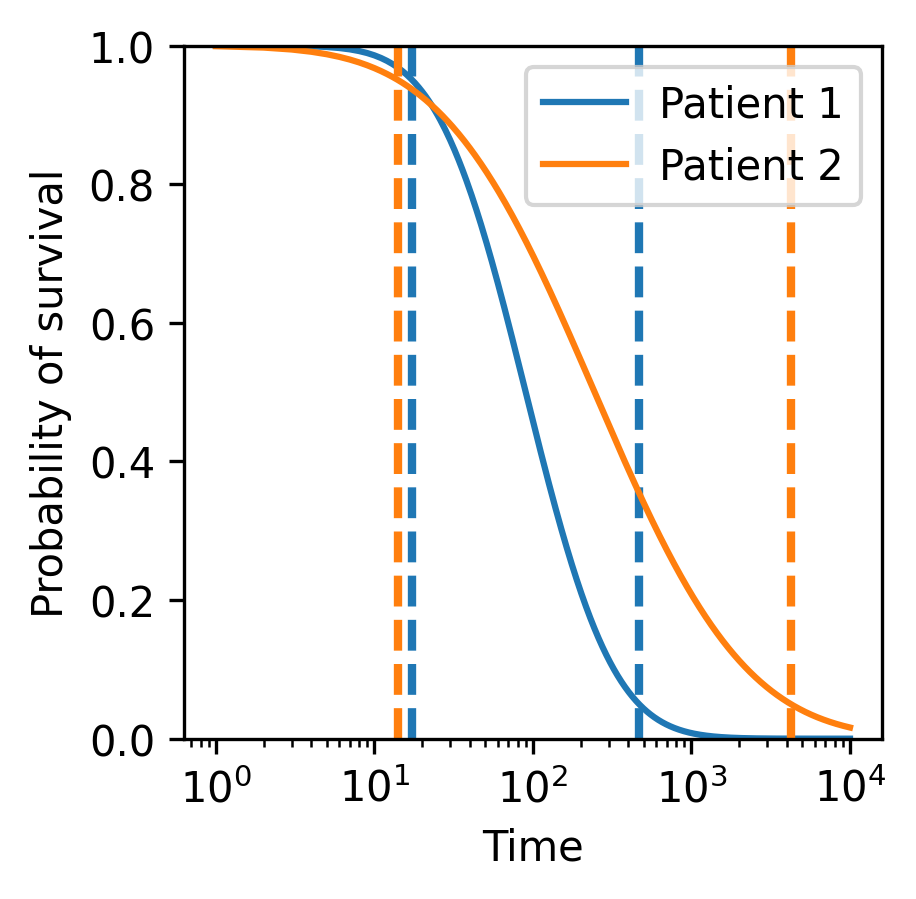}
    \caption{
    Survival function (solid lines) and quantiles $q_{\alpha/2}(x), q_{1-\alpha/2}(x)$ of order $\alpha/2$ and $1-\alpha/2$ of $T\vert X=x \sim \cN(4+0.5 x, \sqrt{x})$ (dashed lines) for two values of $x$: $x=1$ (Patient 1) and $x=3$ (Patient 2). The level is $\alpha=10\%$. The lower predictive bounds $q_{\alpha/2}(x)$ of Patient 1 and Patient 2 are close, whereas the upper predictive bound $q_{1-\alpha/2}(x)$ of Patient 2 differs from that of Patient 1 by an order of magnitude. }
    \label{fig:motivationex}
\end{figure}

Following \cite{candes2023conformalized}, most works have focused on the LPB construction \citep{gui2024conformalized, sesia2024doubly}. This work tackles the lack of an UPB.
As a motivating example, Figure \ref{fig:motivationex} displays the survival functions of two patients $x_1 \neq x_2$ ('Patient 1' and 'Patient 2') along with the prediction intervals $[q_{\alpha/2}(x_i), q_{1-\alpha/2}(x_i)]$, with $q_\beta(x)$ the theoretical $\beta$ quantile of $T \vert X=x$. 
While the survival profiles of the two patients diverge in the long term, they are close in the short-term, resulting in similar LPBs. Hence, 
the combination of the LPB with the UPB allows to built a more complete picture of the survival profile of a patient. 

In this paper, we propose a conformal procedure that constructs two-sided or one-sided prediction sets for the survival time of patients.
Our approach starts from the basic observation that true survival times are observed for the subset of individuals that have not been censored. Hence, the method looks at the conformity of each test point with respect to non-censored individuals and if it is confident in the conformity being large enough, gives a two-sided prediction interval. Otherwise, the data point is considered too uncertain for two-sided prediction and an LPB is returned. 
The main ingredient is the use of conformalized classification to classify the event status (censored or not) of a new patient with large enough confidence. 
The proposed method is model-free, wrapping around any survival model
while controlling the error in finite samples, without making any distributional assumptions besides that the sampled data points are i.i.d. 
In particular, our method does not rely on any assumptions on the censoring mechanism. This is a crucial feature as censoring usually comes from many sources.


Finally, we mention a recent\footnotemark  work by \cite{yi2025survival}, which concurrently proposed a method for generating two-sided prediction intervals for survival times. Their method consists in calibrating a conformal prediction set with the survival times of non-censored individuals, using weighting to adjust for the distribution shift that results from using only non-censored cases. These weights are based on the c.d.f. of the conditional distributions of the censoring, which the authors propose to estimate using a local Kaplan-Meier method. They provide a bound on coverage deviation based on the accuracy of this estimation. By contrast, our guarantee does not rely on accurate estimation of the censoring.

\footnotetext{\cite{yi2025survival} appeared after the first version of our manuscript was deposited on \texttt{arxiv}; our work was conducted independently.}


\section{Two-sided prediction intervals} \label{sec:method}

While a valid LPB can be obtained using conventional CP on the censored survival times, 
constructing a valid non-trivial UPB is not possible without further distributional assumptions. For instance, if $C=c_0$ with $c_0$ some constant, we lack any information about $T \vert T \geq c_0$, and so the only UPB that is valid for all distributions of $T$ is $+\infty$.
However, this does not rule out the possibility of constructing a finite distribution-free UPB for a \textit{subset} of test points. 
Specifically, since a sample of observations $(X_i, \tilde T_i)_{i, \Delta_i=1} = (X_i, T_i)_{i, \Delta_i=1}$ from $P_{X,T \vert \Delta=1}$ is available,  CP can be used to compute a valid two-sided prediction interval for $T_{n+1}$ conditionally on $\Delta_{n+1}=1$. Hence, the basic idea of our approach is as follows. First, a classification rule $\hat \Delta (X)$ of $\Delta$ is learned. If $\hat \Delta(X_{n+1})=1$ with enough confidence, then the procedure returns a \textit{two-sided} prediction interval (PI), i.e. both LPB and UPB. Otherwise, the procedure only returns an LPB. 
Our procedure uses CP at both steps (classification and generation of PIs) to ensure that the error of the method is controlled. A brief introduction to CP is provided in Appendix \ref{app:classicalCP}.

In words, our approach boils down to learning if a new patient $X_{n+1}$ is "similar" to the non-censored patients $(X_i)_{i, \Delta_i=1}$, for which survival times are observed. This way, true survival times can be leveraged to make valid two-sided predictions. 


Let $\hat \Delta_{n+1} = \hat \Delta(X_{n+1}) \in \{0,1\}$ be a decision rule for $\Delta_{n+1}$, $\hat C_1(X_{n+1})$ a prediction interval for $T_{n+1}$, and $\widehat{\text{LB}}_0(X_{n+1})$ an LPB for $T_{n+1}$. Define the prediction set:
\begin{align} \label{eq:twosidedpredset}
\hat C(X_{n+1}) = \hat C_1(X_{n+1}) \hat \Delta_{n+1}+ [\widehat{\text{LB}}_0(X_{n+1}), +\infty[ (1-\hat \Delta_{n+1}). 
\end{align}
Given a nominal level $\alpha \in ]0,1[$, we consider the aim of constructing $\hat \Delta_{n+1}, \hat C_1(X_{n+1}), \widehat{\text{LB}}_0(X_{n+1})$ such that the resulting prediction set $\hat C(X_{n+1})$ contains $T_{n+1}$ with probability above $1-\alpha$: 
\begin{align} \label{eq:error}
\P( T_{n+1} \notin \hat C(X_{n+1})) 
=  \P( T_{n+1} \notin \hat C_1(X_{n+1}), \hat \Delta_{n+1}=1) +  \P( T_{n+1} \leq \widehat{\text{LB}}_0(X_{n+1}), \hat \Delta_{n+1}=0) \leq \alpha. 
\end{align}


\begin{lemma} \label{lem:errorbound}
The coverage error \eqref{eq:error} is bounded as 
\begin{align} 
\P( T_{n+1} \notin \hat C(X_{n+1})) & \leq \P( T_{n+1} \notin \hat C_1(X_{n+1}) \vert \Delta_{n+1}=1)\P( \Delta_{n+1}=1)  + \P(T_{n+1} \leq \widehat{\text{LB}}_0(X_{n+1})) \nonumber\\
&\P( \hat \Delta_{n+1} \neq   
\Delta_{n+1} \vert  \Delta_{n+1}=0)\P( \Delta_{n+1}=0)
\nonumber
\end{align}
\end{lemma}


Lemma \ref{lem:errorbound} implies that to have $\P( T_{n+1} \notin \hat C(X_{n+1})) \leq \alpha$, it suffices to construct the classification rule $\hat \Delta_{n+1}$, the prediction set $ \hat C_1(X_{n+1})$ and the lower predictive bound $\widehat{\text{LB}}_0(X_{n+1})$ such that
    \begin{align}
    \P( \hat \Delta_{n+1} \neq \Delta_{n+1} \vert  \Delta_{n+1}=0) \leq \alpha_1 \label{eq:coverage_classif} \\
        \P( T_{n+1} \notin \hat C_1(X_{n+1}) \vert \Delta_{n+1}=1) \leq \alpha_1 \label{eq:coverage_twosided} \\
        \P(T_{n+1} \leq \widehat{\text{LB}}_0(X_{n+1})) \leq \alpha_0 \label{eq:coverage_onesided}
    \end{align}
   with $\alpha_0 +\alpha_1 = \alpha$. In the sequel, we consider without loss of generality that $\alpha_0 =\alpha_1 = \alpha/2$. We next outline our proposed method to achieve \eqref{eq:coverage_classif}, \eqref{eq:coverage_twosided} and \eqref{eq:coverage_onesided}.

Let $\Dtrain$ and $\Dcal$ be a random splitting of $\{i=1, \dots, n\}$. 
Define $\Dcal^0 = \{ i \in \Dcal \colon \Delta_i=0\}$ and $\Dcal^1 = \{ i \in \Dcal \colon \Delta_i=1\}$. Each of the components of $\hat C(X_{n+1})$ is constructed as follows: 

\begin{itemize}
\item Classification rule $\hat \Delta_{n+1}$: For  $\hat \Delta_{n+1}$, we simply need a version of CP for classification that controls the type 1 error instead of the coverage probability. Formally, such a binary classification rule is a test controlling the type I error at the level $\alpha$ for the null hypothesis $\cH_0 \colon \Delta_{n+1}=0$. A valid test can be constructed based on the \textit{conformal $p$-value}
\begin{align} \label{eq:confpvalue}
p_{n+1}= (1 + \vert \Dcal^0 \vert)^{-1} \left(1+ \sum_{i \in \Dcal^0} \ind \{ \nu_\Delta(X_i, 0) \leq \nu_\Delta(X_{n+1}, 0) \} \right) 
\end{align}
where $\nu_\Delta \colon (x, y) \mapsto \R $ is a non-conformity score function depending only on $\Dtrain$. 
Under the null hypothesis $\cH_0$, $p_{n+1} \sim \cU \{1/{n+1}, \dots, 1 \}$ and so yields a valid test. Hence, the proposed classification rule is
\begin{align} \label{eq:classifrule}
\hat \Delta_{n+1} = \ind \{ p_{n+1} < \alpha/2 \} . 
\end{align}
In the case of classical conformal binary classification, the optimal score function is $(x,y) \mapsto 1- \P(Y=y\vert X=x)$ \citep{sadinle2019least, romano2020classification}. 
Hence, a natural choice for $\nu_\Delta$ is $\nu_\Delta(x,y) = 1 - \hat \pi_y(x)$ with $\hat \pi_y(x)$ an estimate of $\P(\Delta=y\vert X =x)$ returned by a classifier trained on $(X_i, \Delta_i)_{i \in \Dtrain}$.

\item Prediction set $\hat C_1(X_{n+1})$: Conditionally on $\Delta_{n+1}=1$, $(X_{n+1}, T_{n+1})$ together with the non-censored observations $(X_i, \tilde T_i)_{i \in \Dcal^1} = (X_i, T_i)_{i \in \Dcal^1}$ form an exchangeable sequence. Hence, applying CP at the level $\alpha/2$ using as calibration sample the non-censored observations $(X_i, T_i)_{i \in \Dcal^1}$, outputs a prediction set $\hat C_1(X_{n+1})$ satisfying $\P( T_{n+1} \notin \hat C_1(X_{n+1}) \vert \Delta_{n+1}=1) \leq \alpha/2$.  
Letting $\nu_1(x,y) \mapsto \R $ be a non-conformity score function depending only on $\Dtrain$, the proposed prediction set $\hat C_1(X_{n+1})$ is 
\begin{align} \label{eq:predset1}
\hat C_1(X_{n+1})= \{ t>0, \nu_1(X_{n+1},t) \leq \hat q_1 \} ,
\end{align}
where $\hat q_1$ is the quantile of order $1-\alpha_1 \frac{\vert \Dcal^1 \vert +1}{\vert \Dcal^1 \vert }$ of $\{\nu_1(X_i, \tilde T_i), i \in \Dcal^1 \}$. The choice of $\nu_1$ is discussed below.

\item Lower bound $\widehat{\text{LB}}_0(X_{n+1})$: Any LPB method that has marginal coverage at the level $\alpha/2$, such as the naive LPB, or the LPB proposed by \cite{candes2023conformalized} can be used (in the latter case, the validity is only asymptotical). That is, denoting by $\widehat{\text{LB}}_{\text{PI}, \alpha/2}$ such a lower bound method, we plug it into the lower bound $\widehat{\text{LB}}_0(X_{n+1})$: $\widehat{\text{LB}}_0(X_{n+1}) = \widehat{\text{LB}}_{\text{PI}, \alpha/2}(X_{n+1})$. 
\end{itemize}

The procedure is summarized in Algorithm \ref{algo:proc} in Appendix \ref{app:algo}. 
To conclude with the exposition of the method, we discuss the choice of the non-conformity score function $\nu_1$. In general, the choice of non-conformity score function is quite an important consideration in conformal prediction, since on it depends how informative the prediction sets are (their size), and it also has a bearing on \textit{conditional} coverage (coverage conditional on $X_{n+1}$). In contrast to a standard regression setting, the censoring brings nuances that have to be taken into account carefully. 
To start with, we suggest that here it is desirable to train the non-conformity score on the entire population (as opposed to the non-censored one) for two reasons: 1) integrating with an existing survival model is often convenient 2) the score should be as robust as possible against misclassified data points where $\hat \Delta=1$ but $\Delta=0$.
Next, as in the standard regression case, it is desirable to choose a non-conformity score which takes into account data heteroscedascity so that conditional coverage is approximately ensured; a classical choice is the quantile-based score of \cite{romano2019conformalized}. Alternatively, \cite{chernozhukov2021distributional} proposed a score  closely related to that of \cite{romano2019conformalized}, based on the c.d.f.:
\begin{align} \label{eq:surv_score}
    \nu(x,y) = \vert 1/2 - \hat F_{T\vert X =x}(y) \vert. 
\end{align}
This score is proven to control conditional coverage asymptotically under mild assumptions in \cite{chernozhukov2021distributional}. In survival analysis, predictive models generally directly estimate the c.d.f. of the conditional distribution $T \vert X=x$, making this score particularly suitable. 
With this choice of $\nu_1$, the resulting prediction set is:
\begin{align*}
   \hat C_1(X_{n+1})= [\hat F_{T\vert X =x}^{-1}(1/2-\hat q_1), \hat F_{T\vert X =x}^{-1}(1/2+\hat q_1)]. 
\end{align*}

\begin{remark} \label{rem:extrapolation}
    In practice, the estimated survival functions of some individuals may not be invertible in the smaller values: non-parametric survival techniques, which includes the popular Cox model, do not extrapolate beyond the maximum observed censored outcome, at which point the survival function is not necessarily zero. In other words, we cannot test a possible outcome value $y$ beyond $\max_i \tilde T_i$: if this outcome value is accepted, 
    then the UPB is infinite. 
    The interpretation of an infinite UPB is that the event 'surviving beyond the maximum observed censored time' is included in the prediction set, which is expected to occur naturally in the context of censored data. 
\end{remark}

\begin{remark}
    When fitting the classifier for $\nu_{\Delta}$, a different combination of variables can be used than for the predictive model used in $\nu_1$. 
\end{remark}

\begin{theorem} \label{th:guarantee}
If $(X_i, T_i, C_i)_{i=1}^{n+1}$ is exchangeable, then for the procedure given in Algorithm \ref{algo:proc}, we have that $\P( T_{n+1} \notin \hat C(X_{n+1})) \leq \alpha$.  
\end{theorem}
The proof is immediate from the guarantee of split conformal prediction \citep{vovk2005algorithmic, papadopoulos2002inductive}.

We end this section with a short discussion on the applicability of our method. Existing distribution-free methods with finite-sample guarantees only provide an LPB. 
By contrast, our approach returns both an LPB and an UPB for the points such that $\hat \Delta_{n+1}=1$, otherwise returning a plug-in LPB $\widehat{\text{LB}}_{\text{PI},\alpha/2}$ at a reduced level $\alpha/2$ for the points such that $\hat \Delta_{n+1}=0$. 
As a result, the proposed approach  can be considered more informative on the points classified as $\hat \Delta_{n+1}=1$ in the sense that the prediction interval is two-sided, but less informative on the remaining data points $\hat \Delta_{n+1}=0$, since for those points, the plug-in LPB is applied at a level divided by half ($\alpha/2$) in comparison. While a proper comparison is complicated, 
it is clear that if the proportion of $\hat \Delta_{n+1}=1$ is very small then a fully one-sided method makes more sense.
Hence, there is a trade-off that depends on the resulting rate $\P(\hat \Delta_{n+1} =1)$, which itself depends on two main factors: the censoring proportion $\P(\Delta=0)$ and the difficulty of the testing problem $\cH_0 \colon \Delta_{n+1} =0$ vs. $\cH_1 \colon \Delta_{n+1} =1$. In particular, given that the type 1 error is controlled, it is expected that $\P(\hat \Delta_{n+1} =1)$ cannot be larger than $\P(\Delta=1)$. Hence, our approach is better suited to small to moderate censoring rates; for high censoring rates, we recommend using a one-sided approach. Regarding the classification of $\Delta_{n+1}$, it relies on $P_{X \vert \Delta=0}$ and $P_{X \vert \Delta=1}$ being separable, which occurs if $X$ is a strong predictor of $T$, and being observed to die in the study's period of time $\Delta$ strongly depends on $T$. For instance, it is often the case that individuals that live longer get censored whereas individuals with a short survival time have recorded deaths. In general, however, this is not necessarily consistently the case across the entire dataset, with censoring coming from a mixture of sources. Hence, we suggest assessing the feasibility of the classification problem before applying our method in practice, using, for instance, two-sample testing.


\section{Simulation study} \label{sec:simu}
In this section, we study the performance of our procedure on synthetic data. In contrast to real data, the use of synthetic data allows us to compute coverage errors exactly because we have access to the ground truth. 
We consider $X \sim \cU[0,1]^2$, $\log T \vert X \sim \cN(\beta_0 +\beta X^T, 1)$ with $\beta_0=3$ and $\beta=(3,-2)$, and $C \sim \cU[1, t_0]$ with $t_0$ chosen so that the censoring rate is equal to 30\% and 50\% respectively. We use a sample size of $n \in \{400, 800\}$ with a splitting ratio of 50\% for the train-calibration split and a test size of $100$. 
We apply our method 
using either Random 
Forest or Logistic Regression for the classification step, and with the non-conformity score $\nu_1$ given by \eqref{eq:surv_score} for generating the two-sided PIs. Two techniques are considered to estimate the cdf in \eqref{eq:surv_score}: Cox's proportional hazards model and the AFT model with a Weibull parametrization. 
For the one-sided component of our procedure, we consider a setup where censoring times are not observed for non-censored individuals and use Naive LB with the one-sided version of $\nu_1(x,y)$ as non-conformity score: $\nu_{\text{LB}}(x,y) = 1/2-\hat F_{T \vert X=x}(y)$.
This gives the methods \texttt{TwoSidedCP + Naive LB (LR-Cox)}, \texttt{TwoSidedCP + Naive LB (LR-WeibullAFT)}, \texttt{TwoSidedCP + Naive LB (RF-Cox)} and \texttt{TwoSidedCP + Naive LB (RF-WeibullAFT)}. 
More details concerning the classifiers and survival techniques used can be found in Appendix \ref{app:sec:methods}. 
In addition, we compare our method with \cite{yi2025survival} (called \texttt{SCP} therein),
using the Kaplan-Meier estimator to estimate the censoring distribution, and for the non-conformity score we use the same score $\nu_1$ as in \texttt{TwoSidedCP}. 

To evaluate the methods, coverage is assessed separately for two-sided and one-sided groups, the two-sided (one-sided) group being defined as the set of test points  with a finite (infinite) UPB.
Note here that we have explicitely defined the two-sided group as finite UPBs instead of being \textit{assigned} a two-sided prediction (points classified as $\hat \Delta$=1 in the case of \texttt{TwoSidedCP} and all points in the case of \texttt{SCP}), since infinite UPBs can sometimes be generated when calibrating for two-sided prediction (see Remark \ref{rem:extrapolation} in Section \ref{sec:method}). Moreover this can not only occur for \texttt{TwoSidedCP}, but for \texttt{SCP} as well if estimation of censoring distribution is non-parametric, due to not being able to test outcomes values beyond the maximum observed censored time. 

We only summarize the results here; see Appendix \ref{app:simu} for a full report of the numerical results with $\alpha=10\%$. \texttt{TwoSidedCP} achieves coverage in each group in each experiment. By contrast, \texttt{SCP} can have a higher two-sided proportion, but generally displays a lack of coverage in the two-sided group. As expected, as the sample size increases, there is an improvement in the two-sided prediction proportion for \texttt{TwoSidedCP}, and as the censoring rate increases, two-sided prediction proportion decreases for all methods. 


\section{Application to real-data}

In this section, we apply our procedure on public real datasets from the healthcare domain: WHAS \citep{hosmer2008applied}, Rotterdam \& GBSG \citep{foekens2000urokinase, schumacher1994randomized, royston2013external}, and SUPPORT \citep{knaus1995support}. 
The datasets are summarized in Table \ref{tab:datasets_summary} and further described in Appendix \ref{app:sec:datasets}.

\begin{table}[]
\small
    \centering
    \begin{tabular}{lcc}
         Dataset & Censoring rate & Sample size  \\
         \hline
         WHAS & 58\% & 1310 \\
         Rotterdam\&GBSG & 57\% & 2232 \\
         SUPPORT & 32\% & 8873 \\
         METABRIC & 43\% &1904\\
    \end{tabular}
    \caption{Summary of the datasets considered in the real data study.}
    \label{tab:datasets_summary}
\end{table}

Our method is applied using the same specifications as in the simulation study, except that 
the classification is restricted to Random Forest, and we consider an additional survival technique, the deep-learning-based method \texttt{Deepsurv} \citep{katzman2018deepsurv}. 

\begin{figure}
    \centering
    \includegraphics[width=0.35\linewidth]{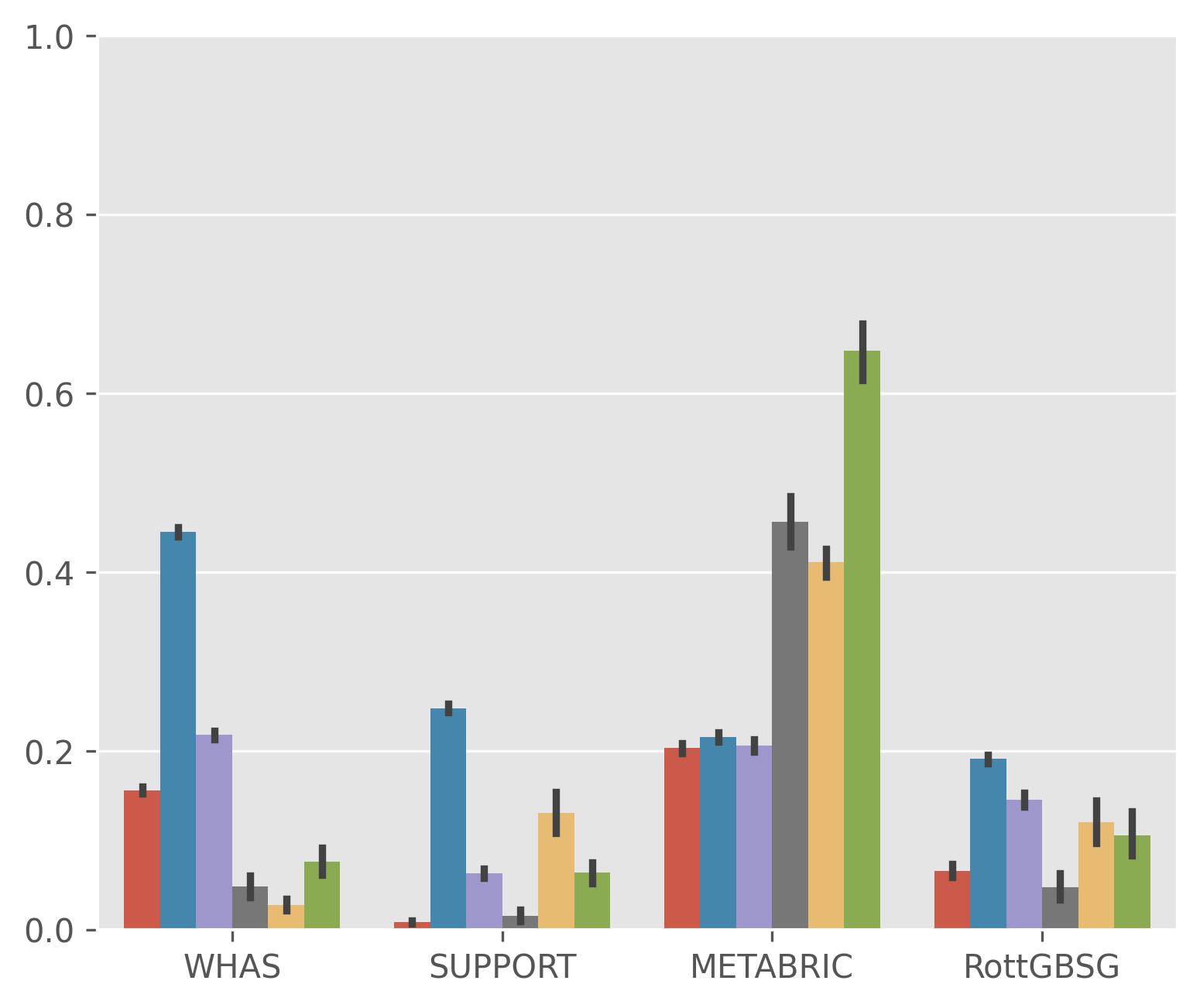}
    \includegraphics[width=0.3\linewidth]{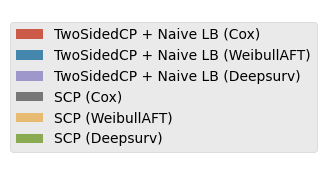}
    \caption{Proportion of finite UPBs generated by each method, in each dataset.}
    \label{fig:2sidedprop}
\end{figure}

\begin{figure}
    \begin{minipage}{0.5\linewidth}
       \centering
\includegraphics[width=\linewidth]{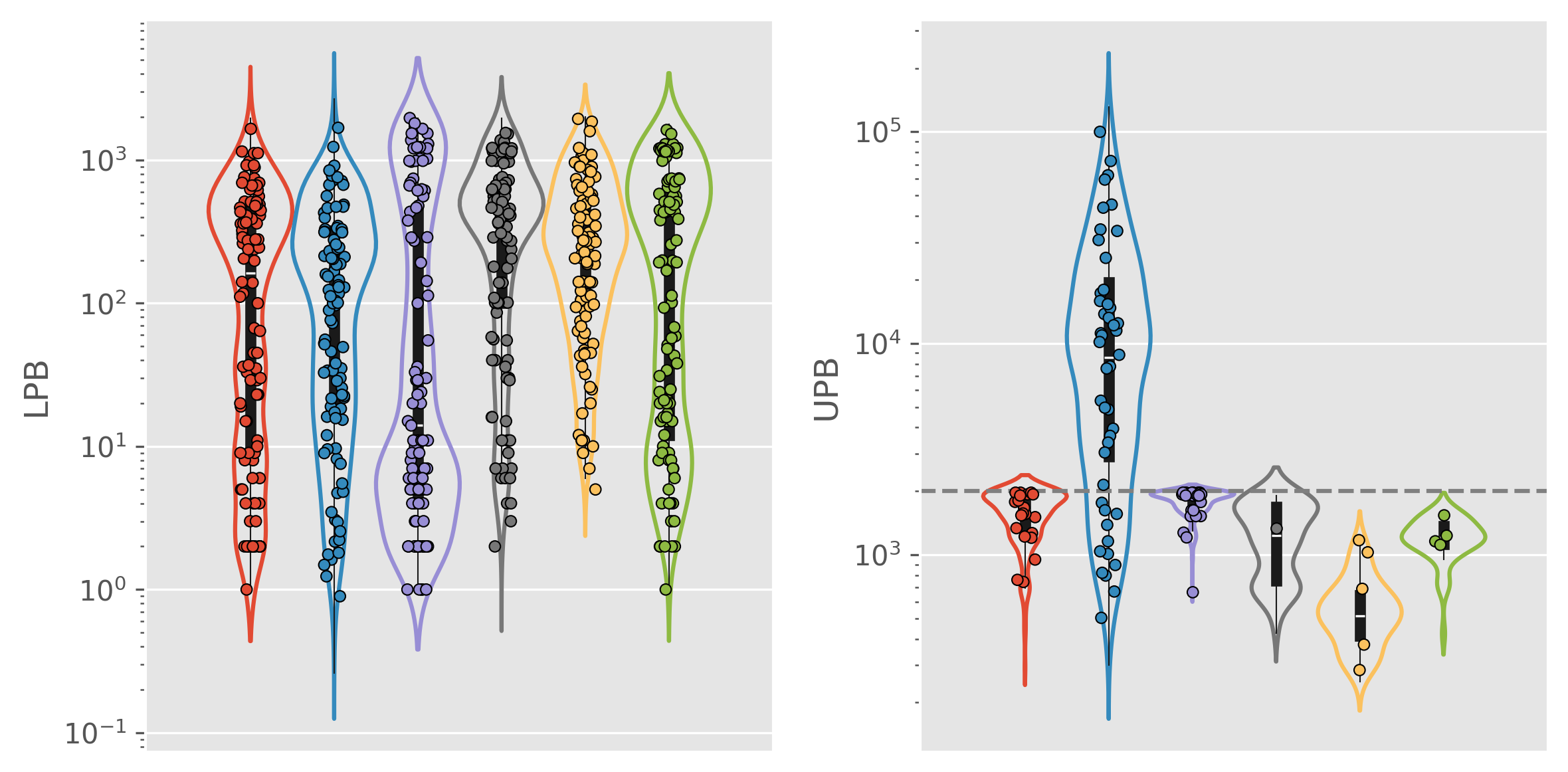}
        \subcaption{WHAS}
    \end{minipage}%
    \begin{minipage}{0.5\linewidth}
       \centering
        \includegraphics[width=\linewidth]{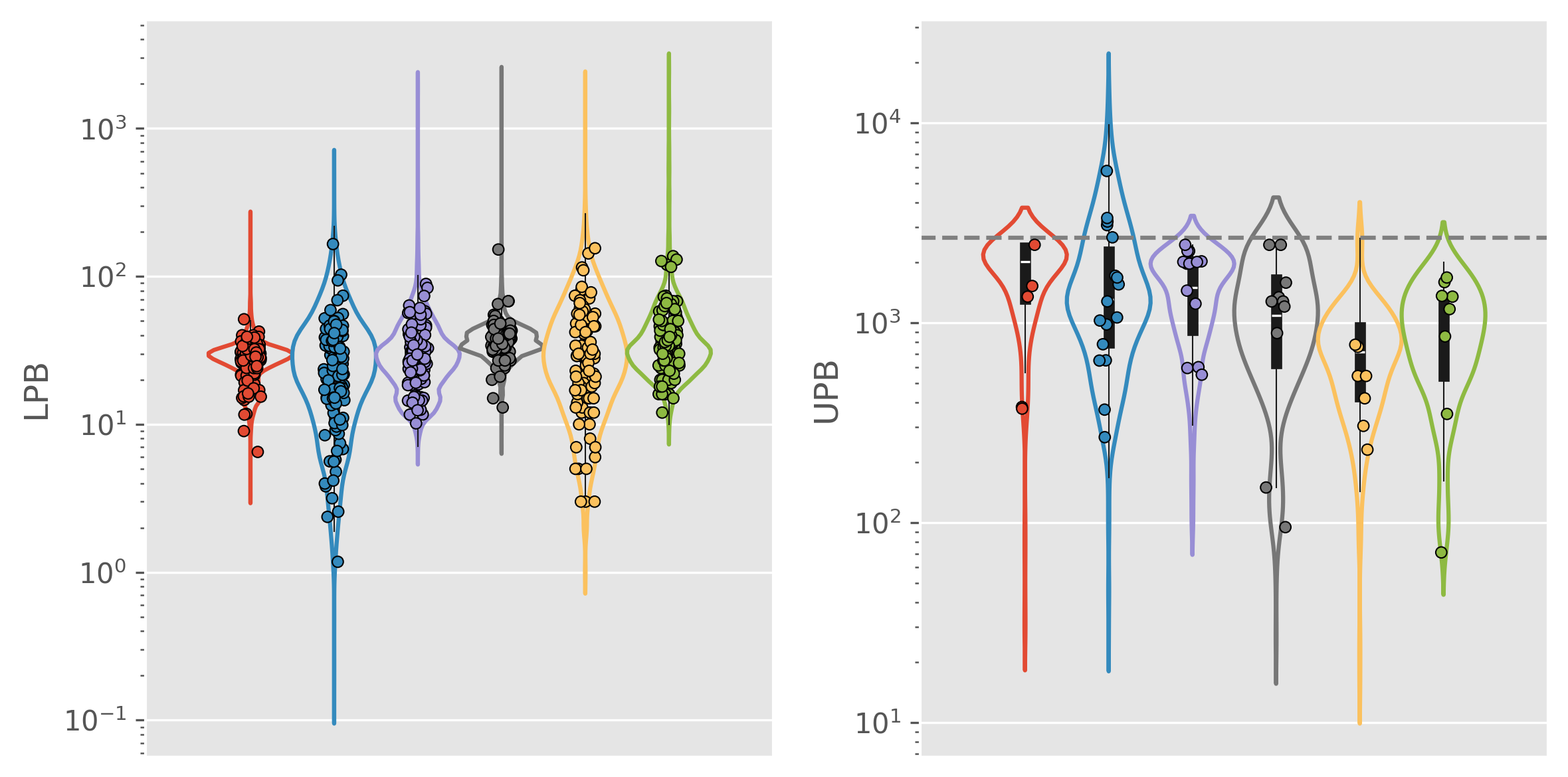}
        \subcaption{Rotterdam\&GBSG}
    \end{minipage}
    \begin{minipage}{0.5\linewidth}
       \centering
        \includegraphics[width=\linewidth]{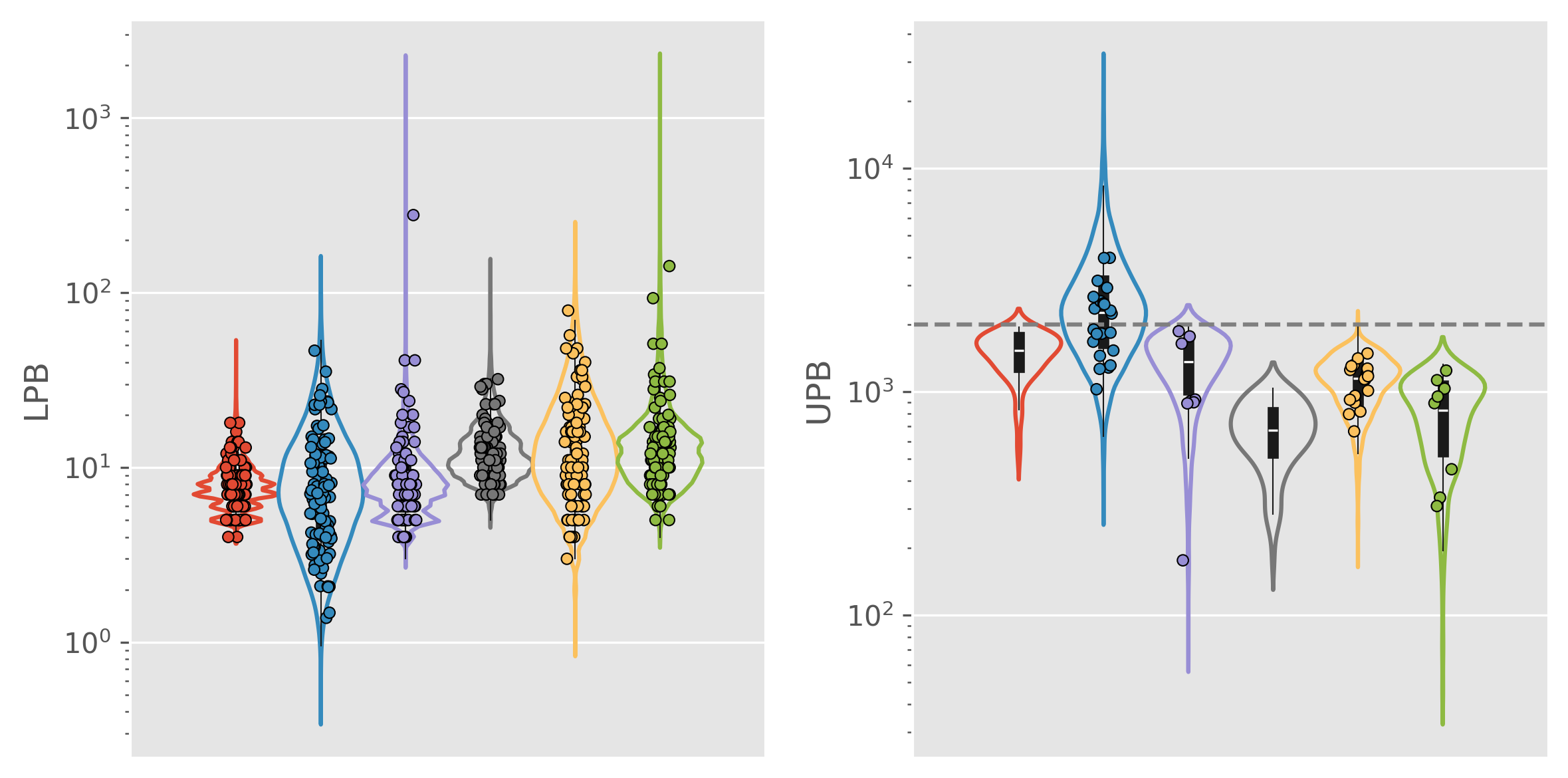}
        \subcaption{SUPPORT}
    \end{minipage}
    \begin{minipage}{0.5\linewidth}
       \centering
        \includegraphics[width=\linewidth]{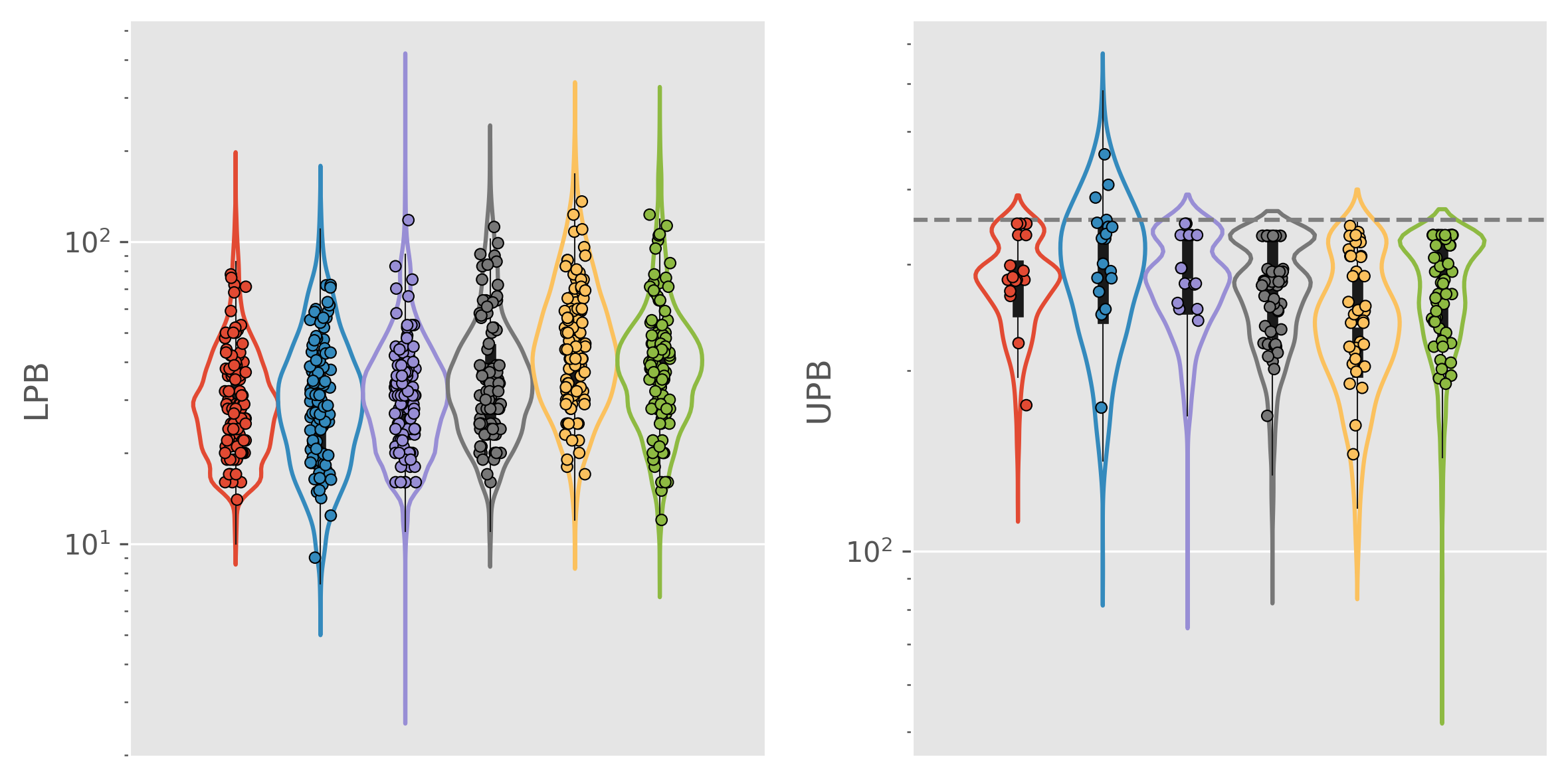}
        \subcaption{METABRIC}
    \end{minipage}
    \includegraphics[width=\linewidth]{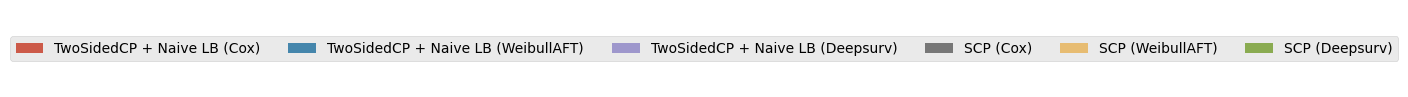}
    \caption{Distribution of predictions generated by each method, pooled across 100 train/calibration/test splits (only finite UPBs are plotted). Points corresponds to the predictions generated for a random sample of 100 patients, in a randomly selected train/calibration/test split. The dashed line represents the maximum observed censored survival time in the dataset. 
    }
    \label{fig:pred}
\end{figure}

For each dataset, we generate 100 random training / calibration / test splits with a splitting ratio of 40\%-40\%-20\%. In each generated test set, since all survival times are not available, the coverage of the generated PIs cannot be computed exactly. Instead, we can compute the following lower bound $\text{cov}_{\text{lo}}$ and upper bound $\text{cov}_{\text{up}}$:
\begin{align*}
    \text{cov}_{\text{lo}} &= 1 - \bigg( \P(T_{n+1} \notin \hat C(X_{n+1}), \Delta_{n+1}=1) + \P(\Delta_{n+1}=0, \hat \Delta_{n+1}=1) \\
    &+ \P(\tilde T_{n+1} \leq \widehat{\text{LB}}_0(X_{n+1}), \Delta_{n+1}=0, \hat \Delta_{n+1}=0) \bigg) \\ 
    \text{cov}_{\text{up}} &= 1-\left(\P(T_{n+1} \notin \hat C(X_{n+1}), \Delta_{n+1}=1) + \P(\tilde T_{n+1} \geq \sup \hat C_1(X_{n+1}),\Delta_{n+1}=0, \hat \Delta_{n+1}=1)\right). 
\end{align*}
However, these bounds are expected to be loose. 
Table \ref{tab:results} displays the (sample) values of $\text{cov}_{\text{lo}}$ and $\text{cov}_{\text{up}}$ in each group
for $\alpha=20\%$, and Figure \ref{fig:2sidedprop} displays the proportion of test points with a finite UPB. 
Our method generally achieves coverage in each group, with $\text{cov}_{\text{lo}}$ either above $\alpha$ or $[\text{cov}_{\text{lo}},\text{cov}_{\text{up}}]$ closely centered on $\alpha$. In terms of the proportion of generated two-sided PIs, our method is competitive with \cite{yi2025survival}: results vary across dataset, with neither method outperforming the other consistently. In 2 out of the 4 datasets, \texttt{WHAS} and \texttt{SUPPORT}, there is a big gap between the proportion of points classified as $\hat \Delta=1$ (equal to the two-sided proportion of \texttt{TwoSidedCP + Naive LB (WeibullAFT)} since it uses a parametric survival model) and the two-sided proportion when using non-parametric survival model; that could be due to a poor calibration of the predictive models, leading to large uncertainties. In those same cases the performance of \texttt{SCP} is also the poorest. In the \texttt{METABRIC} dataset, it has a large two-sided prediction rate, although whether coverage is achieved is unknown. Lastly, all methods perform poorly in the case of \texttt{RottGBSG}; concerning \texttt{TwoSidedCP} we interpret the low rate of $\hat \Delta=1$ to having combined two datasets with different follow-up periods. 
Finally,  Figure \ref{fig:pred} illustrates the predictions generated by each method. The use of a parametric survival model in \texttt{TwoSidedCP} extrapolates in many instances, and more flexible survival models tend to generate PIs that are slightly smaller, with higher LPBs and lower UPBs -- although this is mostly evident concerning the LPBs. 

\begin{table}
    \centering
    \small
    \begin{tabular}{llccccc}
            Dataset  & Method & \multicolumn{2}{c}{One-sided}  & \multicolumn{2}{c}{Two-sided}  \\
         \cline{3-6}
         & & $\text{cov}_{\text{lo}}$ & $\text{cov}_{\text{up}}$ & $\text{cov}_{\text{lo}}$ & $\text{cov}_{\text{up}}$  &\\
         \hline
         \hline
         \multirow{4}{*}{\texttt{WHAS}} & \texttt{TwoSidedCP + Naive LB (Cox)} & 0.93 (0.02) & 0.97 (0.01) & 0.88 (0.05) & 0.88 (0.05) \\
&\texttt{TwoSidedCP + Naive LB (WeibullAFT)} & 0.94 (0.02) & 0.99 (0.01) & 0.79 (0.04) & 0.92 (0.03) \\
         &\texttt{TwoSidedCP + Naive LB (Deepsurv)} & 0.87 (0.03) & 0.98 (0.01) & 0.89 (0.05) & 0.91 (0.05) \\
         \cline{2-7}
         &\texttt{SCP (Cox)} & 0.83 (0.04) & 0.92 (0.02) & 0.76 (0.08) & 0.76 (0.08) \\
&\texttt{SCP (WeibullAFT)} & 0.85 (0.05) & 0.92 (0.02) & 0.55 (0.12) & 0.55 (0.12) \\
         & \texttt{SCP (Deepsurv)} & 0.72 (0.07) & 0.94 (0.02) & 0.79 (0.07) & 0.82 (0.06) \\
         \hline
         \multirow{4}{*}{\texttt{Rotterdam\&GBSG}} & \texttt{TwoSidedCP + Naive LB (Cox)} & 0.92 (0.02) & 0.93 (0.01) & 0.69 (0.13) & 0.86 (0.09) \\
&\texttt{TwoSidedCP + Naive LB (WeibullAFT)} & 0.89 (0.02) & 0.92 (0.02) & 0.70 (0.05) & 0.92 (0.03) \\
         & \texttt{TwoSidedCP + Naive LB (Deepsurv)} & 0.91 (0.02) & 0.93 (0.01) & 0.68 (0.06) & 0.89 (0.05) \\
         \cline{2-7}
         &\texttt{SCP (Cox)} & 0.85 (0.03) & 0.86 (0.03) & 0.54 (0.20) & 0.69 (0.19) \\
&\texttt{SCP (WeibullAFT)} & 0.85 (0.06) & 0.90 (0.04) & 0.69 (0.06) & 0.90 (0.04) \\
        & \texttt{SCP (Deepsurv)} & 0.83 (0.03) & 0.88 (0.02) & 0.58 (0.08) & 0.75 (0.09) \\
          \hline
          \multirow{4}{*}{\texttt{SUPPORT}} & \texttt{TwoSidedCP + Naive LB (Cox)} & 0.91 (0.01) & 0.91 (0.01) & 0.85 (0.12) & 0.97 (0.04) \\
&\texttt{TwoSidedCP + Naive LB (WeibullAFT)} & 0.88 (0.01) & 0.88 (0.01) & 0.85 (0.02) & 0.96 (0.01) \\
          &\texttt{TwoSidedCP + Naive LB (Deepsurv)} & 0.90 (0.01) & 0.90 (0.01) & 0.85 (0.05) & 0.92 (0.04) \\
         \cline{2-7}
         &\texttt{SCP (Cox)} & 0.85 (0.01) & 0.85 (0.01) & 0.68 (0.13) & 0.74 (0.14) \\
&\texttt{SCP (WeibullAFT)} & 0.84 (0.01) & 0.84 (0.01) & 0.77 (0.04) & 0.88 (0.03) \\
         &\texttt{SCP (Deepsurv)} & 0.86 (0.01) & 0.86 (0.01) & 0.69 (0.08) & 0.72 (0.08) \\
          \hline
         \multirow{4}{*}{\texttt{METABRIC}} & \texttt{TwoSidedCP + Naive LB (Cox)} & 0.89 (0.02) & 0.92 (0.02) & 0.76 (0.04) & 0.95 (0.03) \\       
&\texttt{TwoSidedCP + Naive LB (WeibullAFT)} & 0.89 (0.02) & 0.91 (0.02) & 0.77 (0.04) & 0.96 (0.02) \\
&\texttt{TwoSidedCP + Naive LB (Deepsurv)} & 0.89 (0.02) & 0.92 (0.02) & 0.76 (0.05) & 0.95 (0.03) \\
\cline{2-7}
&\texttt{SCP (Cox)} & 0.84 (0.06) & 0.89 (0.04) & 0.62 (0.08) & 0.89 (0.05) \\
&\texttt{SCP (WeibullAFT)} & 0.80 (0.05) & 0.85 (0.04) & 0.60 (0.06) & 0.85 (0.05) \\
&\texttt{SCP (Deepsurv)} & 0.79 (0.09) & 0.90 (0.04) & 0.57 (0.06) & 0.87 (0.05) \\
        \hline
    \end{tabular}
    \caption{Sample values of $\text{cov}_{\text{lo}}$ and $\text{cov}_{\text{up}}$ in the two-sided group (group of individuals with finite UPBs) and in the one-sided group (remaining group), averaged over 100 train/calibration/test splits. Here $\alpha=$20\%.}
    \label{tab:results}
\end{table}

\section{Discussion}
This work introduces a new approach that provides two-sided or one-sided prediction intervals for survival times using conformal prediction. In a nutshell, the proposed procedure classifies individuals into non-censored and censored populations, constructing two-sided prediction intervals for those sufficiently similar to non-censored cases and one-sided intervals (lower bound only) otherwise. 
Although two-sided intervals are generated only for a subset of patients, this subset generally corresponds to individuals at higher risk, as the non-censored population often correlates with shorter survival times. 
A limitation is that the proportion of two-sided predictions is constrained by the level of censoring as well as by the level of separability between censored and non-censored cases. While we do not expect our approach to be powerful in all settings, our real data study demonstrate that the method is applicable across a range of real-world cases. Future work could investigate whether in competing events setup, this information can be efficiently utilized in the classification component. 


\section*{Acknowledgements}
A. Marandon acknowledges support from the Turing-Roche partnership.

\bibliographystyle{apalike}
\bibliography{bibli}

\newpage

\appendix

\section{Conformal prediction} \label{app:classicalCP}

\subsection*{} 

We briefly recall conformal prediction in the general case of classification/regression. Consider a covariate variable $X \in \cX$ and an outcome variable $Y \in \cY$, which is either real-valued or categorical. Let $(X_i, Y_i)_{i=1}^n$ be an i.i.d. sample from $P_{X,Y}$ and let $(X_{n+1}, Y_{n+1})$ be a new independent sample from $P_{X, Y}$, for which we observe $X_{n+1}$ and would like to predict the unobserved outcome $Y_{n+1}$. 
The general aim of CP is to build a prediction set $\hat C_\alpha(X_{n+1})$ such that the probability of $Y_{n+1}$ being in the prediction set (\textit{marginal coverage}) is above a pre-fixed confidence level $1-\alpha$: 
\begin{align*}
 \P( Y_{n+1} \in \hat C_\alpha(X_{n+1}) ) \geq 1-\alpha. 
 \end{align*}
where in the above equation, the probability is over $(X_i, Y_i)_{i=1}^{n+1}$. There exists several versions of the CP method; we next describe the so-called Split CP version which is the one most commonly encountered in the literature, see Remark \ref{req:fullcp} for more details. In Split CP, the labeled sample $(X_i, Y_i)_{i=1}^n$ is split into two parts, $\Dtrain$ and $\Dcal$, with $\Dtrain$ being used to train a model that is predictive of the response. Next, CP relies on the use of a non-conformity score function $\nu \colon (x,y) \mapsto \nu(x,y) \in \R$, measuring the inadequacy of $y$ as a prediction for $x$ (based on the pre-trained predictive model). Typical choices are $\nu(x,y) = \vert y -\hat \mu(x) \vert$ in regression, with $\hat \mu(x)$ the estimated prediction at $x$, and 
$\nu(x,y) = 1-\hat \P(Y= y \vert X=x)$ in classification. Given the non-conformity score $\nu$, CP constructs the following prediction set:
\begin{align*} 
\hat C_\alpha(X_{n+1}) = \{ y \in \cY, \nu(X_{n+1}, y) \leq \hat q_{1-\alpha} \}
\end{align*}
where $\hat q_{1-\alpha}$ is the quantile of order $1-\alpha \frac{\vert \Dcal \vert+1 }{\vert \Dcal \vert}$ of $\{ \nu(X_i, Y_i), i \in \Dcal \}$.  
Under the setup where $(X_i, Y_i)_{i=1}^{n+1}$ are i.i.d.,  
it holds that the prediction set $\hat C_\alpha(X_{n+1})$ is \textit{valid}, that is, $\P( Y_{n+1} \in \hat C_\alpha(X_{n+1}) ) \geq 1-\alpha$ \citep{vovk2005algorithmic, papadopoulos2002inductive}. 
While the coverage guarantee holds for any choice of non-conformity score function $\nu$ (that is independent of $\Dcal$ and $(X_{n+1}, Y_{n+1})$), this choice is an important consideration since it has a bearing on how small the prediction sets are. 
 It also has a bearing on \textit{conditional coverage}, which is coverage conditional on each covariate value $x$. In practice, only considering marginal coverage is not sufficient since it doesn't prevent that in certain parts of the feature space, the coverage is below the confidence level $1-\alpha$. 
 While achieving \textit{conditional} coverage exactly is known to be impossible \citep{lei2014distribution, foygel2021limits}, some choices of score provide conditional coverage asymptotically \citep{romano2019conformalized, chernozhukov2021distributional}, by mimicking the geometric form of the optimal (smallest) prediction set in terms of conditional coverage when the distribution is known \cite{lei2014distribution, gupta2022nested}. Alternative approaches exist to target conditional coverage which directly modify the conformal procedure, see e.g. \cite{guan2023localized, ding2024class,gibbs2025conformal}.

\begin{remark} \label{req:fullcp}
    The independence of the scoring function $\nu$ with respect to $\Dcal$ and $(X_{n+1}, Y_{n+1})$ is not a necessary condition: the exchangeability of $(\nu(X_i,Y_i))_{i \in \Dcal \cup \{n+1\} }$ is sufficient for the coverage guarantee to hold. Here, we have described the so-called \textit{split} CP method, which splits the sample $(X_i, Y_i)_{i=1}^n$ into two parts to achieve this exchangeability property under the setup of i.i.d. observations. Alternatively, exchangeability can be achieved by training a score $\nu^y$ on the augmented dataset $((X_1,Y_1),\dots,(X_n,Y_n), (X_{n+1},y))$ for each possible outcome value $y \in \cY$. This second version of CP, called \textit{full} CP, avoids sample splitting at the cost of computational complexity. Moreover, between \textit{split} CP and \textit{full} CP, there exists intermediate approaches, see \cite{barber2021predictive, gupta2022nested}. 
\end{remark}



\section{Algorithm} 
\label{app:algo}

Algorithm \ref{algo:proc} summarizes the procedure proposed in Section \ref{sec:method}. 

\begin{algorithm}
Input: {Level $\alpha$, training sample $(X_i, \tilde T_i, \Delta_i)_{i \in \Dtrain}$, calibration sample $(X_i, \tilde T_i, \Delta_i)_{i \in \Dcal}$, new data point $X_{n+1}$, lower predictive bound function $\widehat{\text{LB}}_{\text{PI}, \alpha/2}(\cdot) $ valid at the level $\alpha/2$.
Define $\Dcal^0 = \{ i \in \Dcal \colon \Delta_i=0\}$ and $\Dcal^1 = \{ i \in \Dcal \colon \Delta_i=1\}$. } \\
1. Learn the non-conformity scores $ \nu_\Delta$ and $ \nu_1$ using $(X_i, \tilde T_i, \Delta_i)_{i \in \Dtrain}$\\
2. Compute the quantile $\hat q^\Delta_{1-\alpha/2}$ of order $1-\alpha/2 \frac{\vert \Dcal^0 \vert+1 }{\vert \Dcal^0 \vert}$ of $\{ \nu_\Delta(X_i, \Delta_i), i \in \Dcal^0\}$ \\
3. Compute the quantile $\hat q^1_{1-\alpha/2}$ of order $1-\alpha/2 \frac{\vert \Dcal^1 \vert +1}{\vert \Dcal^1 \vert }$ of $\{ \nu_1(X_i, \tilde T_i), i \in \Dcal^1\}$ \\
4. For each new test data point $X_{n+1}$
\begin{itemize}
    \item Compute $\hat \Delta_{n+1} = \ind_{ \nu_\Delta(X_{n+1},0) \geq \hat q^\Delta_{1-\alpha/2} }$ 
    \item If $\hat \Delta_{n+1}=1$, compute $\hat C_1(X_{n+1}) = \{t>0,  \nu_1(X_{n+1},t) \leq \hat q^1_{1-\alpha/2} \}$ 
\end{itemize}
5. Compute the prediction set $\hat C(X_{n+1}) = \hat C_1(X_{n+1})$ if $\hat \Delta_{n+1} =1$, $\hat C(X_{n+1}) =[\widehat{\text{LB}}_{\text{PI}, \alpha/2}(X_{n+1}), +\infty[$ otherwise  \\
Output: {Prediction set $\hat C(X_{n+1})$ for each test data point $X_{n+1}$. }
 \caption{Two-sided conformalized survival analysis procedure}\label{algo:proc}
\end{algorithm}

\section{Numerical results for the simulation study} \label{app:simu}
 Tables \ref{tab:simuresults_A} provides numerical results for the simulations described in Section \ref{sec:simu}. 
 
\begin{sidewaystable}
    \centering
    \footnotesize
    \begin{tabular}{cclccccc}
    \toprule
            \multirow{2}{*}{$n$} & \multirow{2}{*}{\makecell{Censoring\\ rate}} & \multirow{2}{*}{Method} & \multirow{2}{*}{\makecell{Two-sided\\ proportion}} &\multicolumn{2}{c}{Coverage}  & \multicolumn{2}{c}{Average length}  \\
         \cmidrule{5-8}
         & & & & One-sided & Two-sided & One-sided & Two-sided\\
         \midrule
          \multirow{12}{*}{400} & \multirow{6}{*}{30\%} & \texttt{TwoSidedCP + Naive LB (RF-Cox)} &0.12 (0.09) &0.98 (0.03) &0.90 (0.07) &5.99 (1.97) &66.12 (23.43) \\
         &&\texttt{TwoSidedCP + Naive LB (RF-WeibullAFT)} &0.12 (0.07)&0.98 (0.02) &0.95 (0.07) &7.19 (2.19) &56.56 (9.62)\\      
         &&\texttt{TwoSidedCP + Naive LB (LR-Cox)} & 0.16 (0.08) & 0.98 (0.02) & 0.91 (0.09) & 5.09 (1.68) & 45.35 (18.22) \\
         &&\texttt{TwoSidedCP + Naive LB (LR-WeibullAFT)} & 0.17 (0.08) & 0.98 (0.02) & 0.94 (0.07) & 7.11 (1.77) & 36.64 (9.91) \\
         &&\texttt{SCP (Cox)} &0.19 (0.25) &0.94 (0.03) &0.85 (0.10) &11.82 (6.78) &70.47 (6.25) \\
         &&\texttt{SCP (WeibullAFT)} &0.18 (0.26) &0.95 (0.04) &0.87 (0.07) &11.17 (6.66)&59.31 (21.01)\\
         \cmidrule{2-8}
 &\multirow{6}{*}{50\%} &\texttt{TwoSidedCP + Naive LB (RF-Cox)} & 0.10 (0.06) & 0.99 (0.01) & 0.90 (0.14) & 3.20 (1.14) & 45.37 (11.14) \\
&&\texttt{TwoSidedCP + Naive LB (RF-WeibullAFT)} & 0.13 (0.06) & 1.00 (0.01) & 0.90 (0.09) & 4.23 (1.16) & 41.39 (11.27) \\
&&\texttt{TwoSidedCP + Naive LB (LR-Cox)} & 0.14 (0.07) & 0.99 (0.01) & 0.89 (0.10) & 3.16 (1.23) & 40.52 (12.71) \\
&&\texttt{TwoSidedCP + Naive LB (LR-WeibullAFT)} & 0.16 (0.06) & 1.00 (0.01) & 0.91 (0.08) & 4.11 (1.32) & 30.41 (7.88) \\
&&\texttt{SCP (Cox)} & 0.02 (0.08) & 0.93 (0.03) & 0.65 (0.27) & 8.46 (2.27) & 35.55 (10.98) \\
&&\texttt{SCP (WeibullAFT)} & 0.02 (0.09) & 0.94 (0.03) & 0.75 (0.05 & 9.63 (2.52) & 36.72 (6.91) \\
\midrule
 \multirow{12}{*}{800} & \multirow{6}{*}{30\%} & \texttt{TwoSidedCP + Naive LB (LR-Cox)} & 0.20 (0.07) & 0.99 (0.01) & 0.88 (0.08) & 5.48 (1.20) & 43.00 (11.36) \\
&&\texttt{TwoSidedCP + Naive LB (LR-WeibullAFT)} & 0.19 (0.06) & 0.98 (0.02) & 0.91 (0.07) & 7.33 (1.38) & 34.93 (6.26) \\
&&\texttt{TwoSidedCP + Naive LB (RF-Cox)} & 0.14 (0.07) & 0.98 (0.02) & 0.88 (0.11) & 5.19 (1.20) & 55.53 (17.29) \\
&&\texttt{TwoSidedCP + Naive LB (RF-WeibullAFT)} & 0.16 (0.06) & 0.98 (0.02) & 0.92 (0.07) & 7.07 (1.40) & 48.89 (10.86) \\
&&\texttt{SCP (Cox)} & 0.25 (0.27) & 0.95 (0.04) & 0.81 (0.06) & 9.96 (3.70) & 63.45 (15.29) \\
&&\texttt{SCP (WeibullAFT)} & 0.28 (0.29) & 0.95 (0.03) & 0.85 (0.05) & 12.93 (4.38) & 64.05 (12.16) \\
\cmidrule{2-8}
&\multirow{6}{*}{50\%} & \texttt{TwoSidedCP + Naive LB (LR-Cox)} & 0.17 (0.05) & 0.99 (0.01) & 0.88 (0.08) & 3.73 (0.79) & 36.04 (8.14) \\
&&\texttt{TwoSidedCP + Naive LB (LR-WeibullAFT)} & 0.18 (0.05) & 1.00 (0.01) & 0.90 (0.08) & 4.49 (1.06) & 29.89 (4.63) \\
&&\texttt{TwoSidedCP + Naive LB (RF-Cox)} & 0.13 (0.05) & 0.99 (0.01) & 0.89 (0.10) & 3.53 (0.79) & 39.79 (9.32) \\
&&\texttt{TwoSidedCP + Naive LB (RF-WeibullAFT)} & 0.14 (0.05) & 1.00 (0.01) & 0.91 (0.09) & 4.33 (0.92) & 38.54 (8.50) \\
&&\texttt{SCP (Cox)} & 0.11 (0.15) & 0.95 (0.03) & 0.75 (0.13) & 9.48 (2.82) & 34.36 (7.18) \\
&&\texttt{SCP (WeibullAFT)} & 0.14 (0.17) & 0.95 (0.03) & 0.81 (0.08) & 11.21 (3.48) & 36.03 (5.79) \\
\bottomrule
    \end{tabular}
    \caption{Results for the simulations described in  Section \ref{sec:simu}: empirical coverage, proportion of finite UPBs ('Two-sided proportion') and average length (average LPB value in one-sided group and average PI size in two-sided group) for each method, using 100 simulations repetitions. Here, $\alpha=$10\%, $n_{\text{train}}=n/2, n_{\text{cal}}=n/2, n_{\text{test}}=100$.}
    \label{tab:simuresults_A}
\end{sidewaystable}

\section{Additional details for the numerical experiments}

\subsection{Datasets used in the real data study} \label{app:sec:datasets}
\begin{figure}
    \centering
    \begin{minipage}[b]{0.2\linewidth}
        \includegraphics[width=\linewidth]{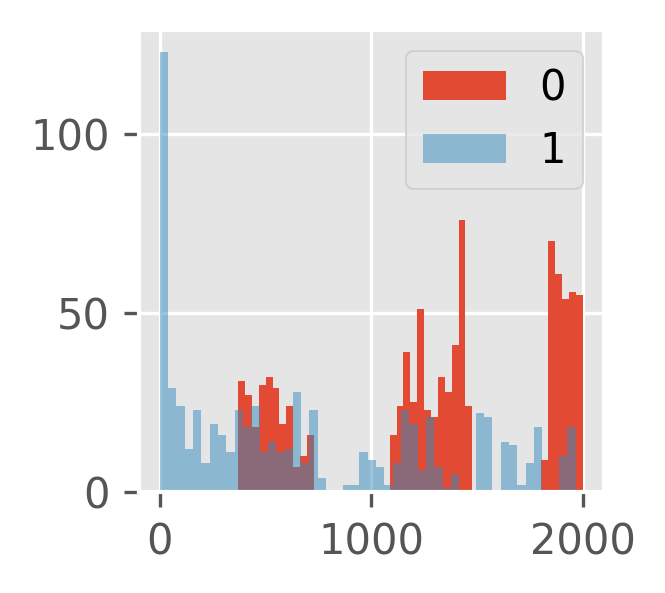}
        \subcaption{WHAS}
    \end{minipage}%
    \begin{minipage}[b]{0.2\linewidth}
        \includegraphics[width=0.95\linewidth]{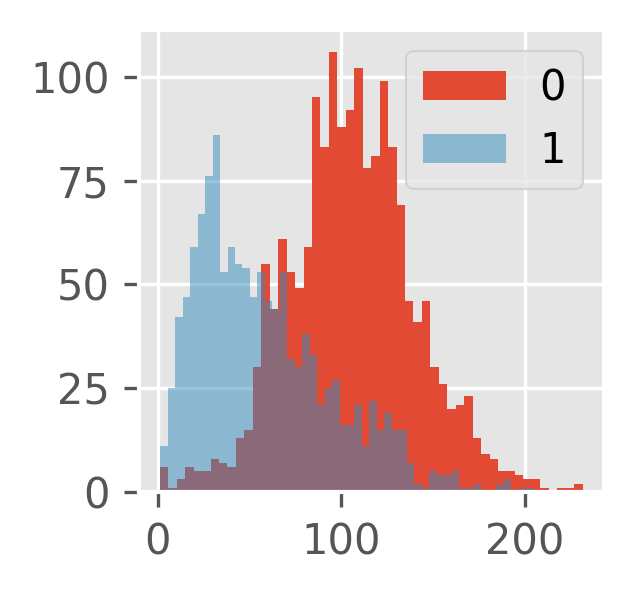}
        \subcaption{Rotterdam}
    \end{minipage}%
    \begin{minipage}[b]{0.2\linewidth}
        \includegraphics[width=0.9\linewidth]{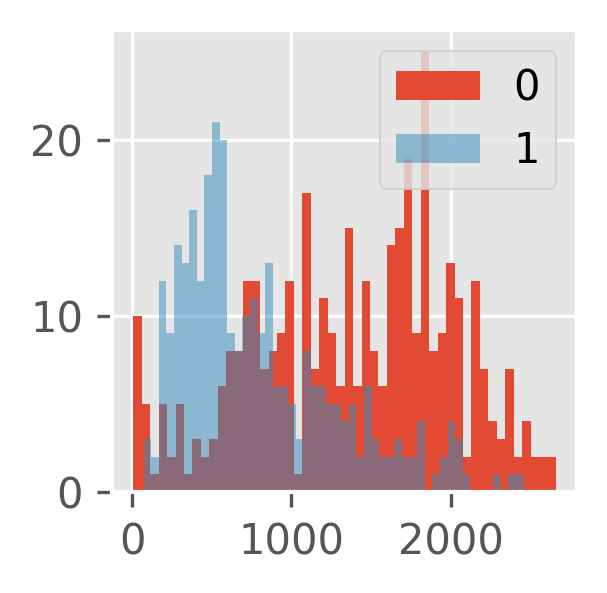}
        \subcaption{GBSG}
    \end{minipage}%
    \begin{minipage}[b]{0.2\linewidth}
        \includegraphics[width=\linewidth]{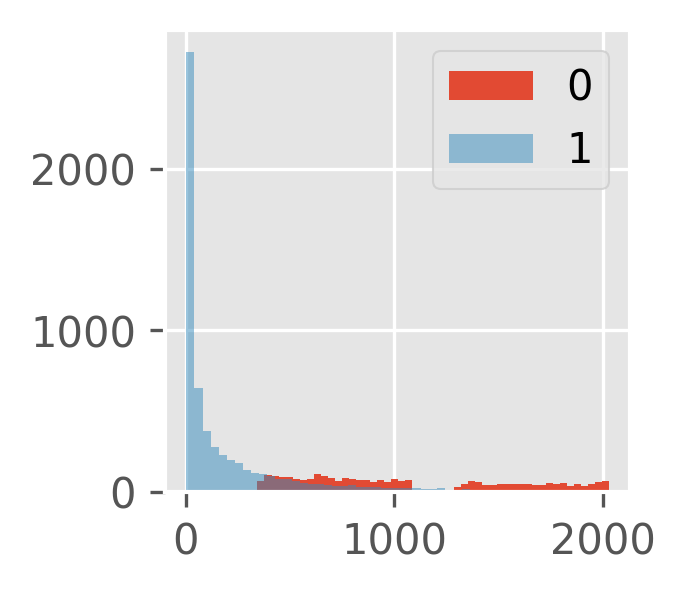}
        \subcaption{SUPPORT}
    \end{minipage}%
    \begin{minipage}[b]{0.2\linewidth}
        \includegraphics[width=0.9\linewidth]{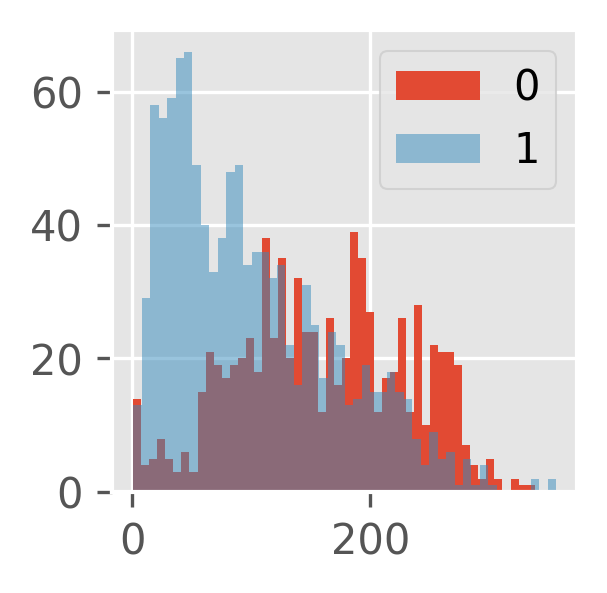}
        \subcaption{METABRIC}
    \end{minipage}%
    \caption{Distribution of the censored survival time for censored samples ('0') and non-censored samples ('1') in each dataset.}
    \label{app:fig:distrib}
\end{figure}

Each dataset is described below. In addition, Figure \ref{app:fig:distrib} shows the distribution of the censored survival time in each dataset. 

\begin{itemize}
    \item \textbf{WHAS} \citep{hosmer08} is a dataset recording survival times of 1310 individuals following 
hospital admission for acute myocardial infarction. There are 6 features consisting of sex, age, body mass index, left heart failure complications and order of myocardial infarction. By the end of the WHAS study, 42.1 percent of the subjects had died of acute myocardial infarction.
    \item  \textbf{Rotterdam \& GBSG} \citep{foekens2000urokinase, schumacher1994randomized} combines two datasets recording recurrence-free survival times of individuals with primary breast cancer. The recurrence-free survival time is defined as the time from primary surgery to the earlier of disease recurrence or death from any cause. \cite{royston2013external} used the Rotterdam data to create a fitted survival model, and the GBSG data for validation of the model. We follow the pre-processing steps of  \cite{royston2013external} to combine the two datasets. After pre-processing, there are 2232 individuals and 7 features consisting of age, number of positive lymph nodes, tumour size, progesterone receptors, oestrogen receptors, menopausal status and hormonal treatment. Death or recurrence is observed for 43\% of the patients. 
    \item  \textbf{SUPPORT} \citep{knaus1995support} is a dataset recording the survival time of critically ill individuals following hospital admission. Patients are combined from 2 studies, each of which lasted 2 years (1989-1991 and  1992-1994). Patients were followed for six months after inclusion in the study, and those who did not die within six months or were lost to follow-up were matched against the National Death Index to identify deaths through 1997. More information can be found at \url{https://archive.ics.uci.edu/dataset/880/support2}. We follow the pre-processing of \cite{katzman2018deepsurv}, selecting 14 features for which almost all patients have observed entries (age, sex, race, number of comorbidities, presence of diabetes, presence of dementia, presence of cancer, mean arterial blood pressure, heart rate, respiration rate, temperature, white blood cell count, serum’s sodium, and serum’s creatinine) and dropping patients with any missing features. The resulting dataset consists of 8873 patients, from which a total of 68.10 percent have a recorded death time.
    \item \textbf{METABRIC} \citep{curtis2012genomic} is a dataset recording the survival time of 1,980 breast cancer patients, from which 57.72 percent have an observed death due to breast cancer. The dataset contains gene expression data and clinical features, and we follow the processing of \cite{katzman2018deepsurv}, selecting 4 genes indicators (MKI67, EGFR, PGR, and ERBB2) and the patient's clinical features(hormone treatment indicator, radiotherapy indicator, chemotherapy indicator, ER-positive indicator, age at diagnosis) to predict survival.
\end{itemize}

\subsection{Implementation details for the classification / survival methods used in the synthetic and real data studies} \label{app:sec:methods}
\begin{itemize}
    \item Random Forest is used with 1000 estimators and we tune the minimum number of samples required to split a node (respectively to generate a leaf node) using 5-fold cross-validation over the grid $[2,10]$ (respectively $[1,5]$). Logistic Regression is used with no regularization penalty and a maximum number of iterations of 100. For both algorithms, we use the Python package \texttt{sckikit-learn} implementation with default parameters otherwise. 
    \item For Cox's model, we use the implementation of the Python package \texttt{scikit-surv} with a combination of $\ell_1$ and $\ell_2$ penalty. The ratio of $\ell_1$ to $\ell_2$ is 90\% and the penalty coefficient is tuned using 5-fold cross-validation and the $c$-index as the score for tuning, with features being standardized. 
    \item For the AFT model with Weibull parametrization, we use the implementation of the Python package \texttt{lifelines} with no penalty. 
    \item \texttt{Deepsurv} is applied using the implementation of the Python package \texttt{pycox} with a 2-layer neural network of 32 neurons in each hidden layer. For training the network, dropout is used with a coefficient of 0.1, batch normalization is used, the optimizer is Adam with a learning rate set to 0.01, the batch size is 256 and the number of epochs is 512. We use early stopping with 20\% of the training set as a validation set. 
\end{itemize}

\end{document}